\documentclass[journal]{IEEEtran}

\usepackage{amsmath,amssymb}
\usepackage{booktabs}
\usepackage{graphicx}
\usepackage{multirow}
\usepackage{array}
\usepackage{url}
\usepackage{xcolor}
\usepackage{cite}

\begin{document}

\title{Patient-Conditioned Dual Hypergraph Reasoning for Auditable Traditional Chinese Medicine Prescription Support}

\author{Weizhi~Nie,
        Shaojin~Bai,
        Weijie~Wang,
        and~Yuting~Su%
\thanks{This work is a computational decision-support study. It is not intended for autonomous clinical prescription.}
\thanks{Weizhi Nie, Shaojin Bai, Weijie Wang, and Yuting Su are with Tianjin University, Tianjin, China.}}

\markboth{IEEE Journal of Biomedical and Health Informatics,~Vol.~XX, No.~X, 2026}%
{Nie \MakeLowercase{\textit{et al.}}: Patient-Conditioned Dual Hypergraph Reasoning for TCM Prescription Support}

\maketitle

\begin{figure*}[!t]
  \centering
  \includegraphics[width=0.96\textwidth]{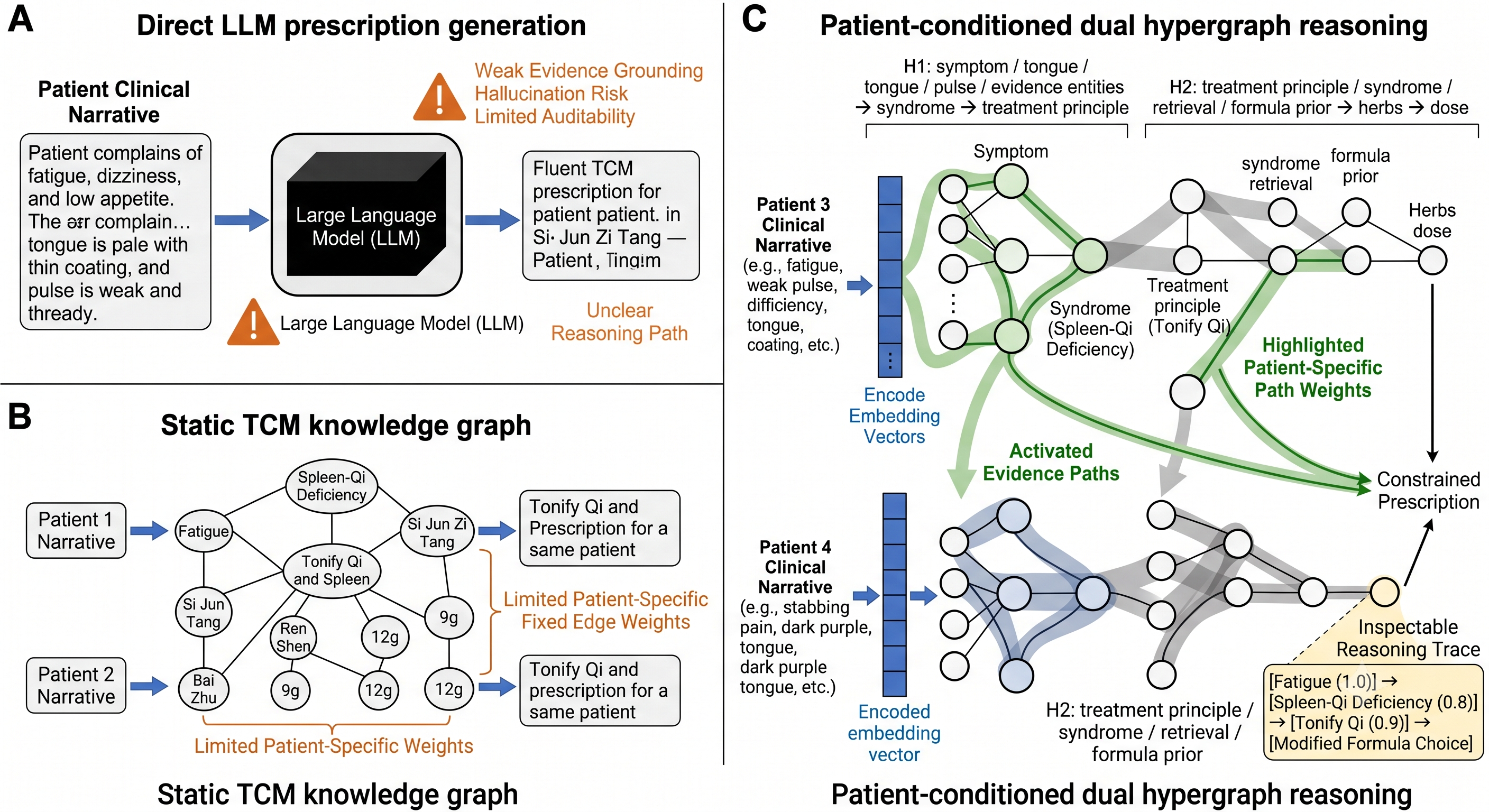}
  \caption{Motivation for patient-conditioned dual hypergraph reasoning in TCM prescription support. Direct language-model prescription generation can be fluent but weakly auditable, while static TCM knowledge resources provide explicit priors but do not adapt their evidential weights to individual patients. The proposed direction activates patient-specific diagnostic and prescription hypergraph paths so that syndrome differentiation, treatment-principle reasoning, herb selection, and dose support remain constrained by inspectable domain knowledge.}
  \label{fig:motivation}
\end{figure*}

\begin{abstract}
Traditional Chinese medicine (TCM) prescription support requires patient-specific reasoning from clinical narratives to syndromes, treatment principles, herbs, and doses. Direct language-model generation can produce fluent prescriptions, but its decisions are difficult to audit against explicit clinical evidence. Static TCM knowledge resources provide useful priors, but they cannot determine which diagnostic and prescription relations should be emphasized for an individual patient. We propose a patient-conditioned dual hypergraph framework for auditable TCM prescription support. The first hypergraph organizes symptom, tongue, pulse, and other clinical evidence around syndrome and treatment-principle reasoning. The second hypergraph organizes syndrome, treatment, disease-context, herb, retrieval, and dose-prior evidence for prescription construction. Unlike static knowledge graphs or fixed hypergraphs, both hypergraphs are dynamically weighted by the patient representation. This design enables individualized activation of diagnostic and prescription paths, supporting personalized syndrome differentiation and herb-dose recommendation while preserving case-level auditability. Experiments on TCM-SD show that dynamic weighting in the first hypergraph improves MacBERT syndrome differentiation to 0.8297 accuracy and 0.3288 macro-F1. On TCM-BEST4SDT, the second hypergraph achieves the best mean Herb-F1 of 0.3111 across three seeds, and the full connected pipeline reaches 0.3074 Herb-F1, close to the oracle setting. A 50-case real-world CAP audit further suggests practical review potential, while highlighting the need for prospective dose-safety validation.
\end{abstract}


\section{Introduction}

\IEEEPARstart{T}{raditional} Chinese medicine (TCM) prescription is a structured clinical reasoning process rather than a single text-generation task. A clinician interprets chief complaints, symptom evolution, tongue and pulse signs, disease context, and treatment response. The clinician then maps this evidence to syndrome differentiation, therapeutic principles, formula organization, herb selection, and dose adjustment. This reasoning chain is relational and many-to-many. Multiple symptoms may jointly support one syndrome, one syndrome may imply several therapeutic principles, and one principle may activate a coordinated group of formula structures and herb roles. These properties make TCM prescription support a useful but difficult setting for interpretable biomedical artificial intelligence.

Recent language models provide useful interfaces for clinical text processing \cite{devlin2019bert,lee2020biobert,alsentzer2019clinicalbert,singhal2023clinical,qwen2024technical}. They can summarize narratives, extract signs, and produce fluent explanations. However, direct generation is poorly aligned with the requirements of medical decision support. A generated prescription may be linguistically plausible while remaining disconnected from explicit syndrome evidence or domain constraints. In biomedical settings, model outputs should therefore be auditable and constrained by prior knowledge rather than treated as free-form natural language decisions \cite{rajkomar2018ehr,li2020behrt}.

Knowledge resources for TCM provide a complementary foundation. Databases such as SymMap, ETCM, TCMSP, TCMID, BATMAN-TCM, and HERB encode relations among symptoms, diseases, syndromes, formulas, herbs, ingredients, targets, and pharmacological evidence \cite{wu2019symmap,xu2019etcm,zhang2023etcmv2,ru2014tcmsp,xue2013tcmid,liu2016batman,fang2021herb}. Yet these resources are typically static. They specify which relations are possible, but they do not directly estimate which relations should be emphasized for a particular patient.

We study TCM prescription support as patient-conditioned activation over structured prior knowledge. Hypergraphs are well suited to this setting because a hyperedge can connect several symptoms to a syndrome, or several herb roles to a formula strategy \cite{zhou2006hypergraph,feng2019hgnn,jiang2019dhgnn}. Unlike pairwise graph encoders \cite{kipf2017gcn,hamilton2017graphsage,schlichtkrull2018rgcn}, hypergraphs can represent higher-order diagnostic and prescription patterns directly. We propose a dual hypergraph framework. H1 models symptom--syndrome--treatment reasoning, H2 models treatment--formula--herb--dose reasoning, and both hypergraphs are dynamically weighted by patient-specific clinical representations.

The central contribution is not an autonomous prescription model. Instead, the framework separates language understanding, patient-conditioned hypergraph weighting, knowledge-constrained inference, and explanation generation. Language representations process patient narratives, while dynamically activated hypergraph paths constrain diagnostic and prescription decisions. The same representation supports prediction, consistency checking, and case-level audit.

The contributions are as follows.
\begin{itemize}
  \item We formulate TCM prescription support as a dual-hypergraph biomedical reasoning problem linking clinical narratives, syndrome differentiation, therapeutic principles, and herb-dose decisions through two coupled higher-order reasoning spaces.
  \item We introduce patient-conditioned dynamic hypergraph weighting for both H1 and H2, where global TCM priors are retained as anchors while patient text representations generate individualized path weights for diagnosis and prescription reasoning.
  \item We evaluate the framework on TCM-SD and TCM-BEST4SDT with fixed-versus-dynamic weighting comparisons, H2 path ablations, direct language-model generation, and end-to-end error propagation, reporting both path-constrained gains and current limitations.
\end{itemize}

\section{Related Work}

\subsection{Clinical Language Models and Biomedical Decision Support}

Transformer language models have become standard components for biomedical and clinical natural language processing. BERT \cite{devlin2019bert}, Chinese BERT with whole-word masking \cite{cui2021wwm}, BioBERT \cite{lee2020biobert}, and ClinicalBERT \cite{alsentzer2019clinicalbert} show the value of pretrained representations for medical text mining. More recent large language models encode broad clinical knowledge \cite{singhal2023clinical} and can support clinical question answering and summarization. However, direct generation raises concerns about hallucination, instability, and limited traceability. These concerns are central to clinical decision support because outputs need evidence, constraints, and reviewability \cite{rajkomar2018ehr,li2020behrt}. Our work uses language models as structured components rather than as unconstrained prescribers.

Recent clinical AI studies have explored temporal attention, causal inference, and multimodal priors for intensive-care analysis, sepsis detection, surgical diagnosis, chest X-ray diagnosis, and thoracic disease classification \cite{nie2023tscan,li2024cisepsis,zhang2025cabg,liu2026causalcompnet,li2023thoracic}. These studies reflect a broader move toward constrained and clinically reviewable modelling. They do not, however, address the symptom--syndrome--treatment--herb reasoning chain required by TCM prescription support.

\subsection{Knowledge Resources for Traditional Chinese Medicine}

TCM informatics has benefited from large knowledge bases linking herbs, formulas, compounds, targets, symptoms, and diseases. SymMap connects TCM concepts with symptom mapping \cite{wu2019symmap}. ETCM provides formula and herb annotations \cite{xu2019etcm,zhang2023etcmv2}. TCMSP, TCMID, BATMAN-TCM, and HERB provide complementary molecular, pharmacological, and literature-guided resources \cite{ru2014tcmsp,xue2013tcmid,liu2016batman,fang2021herb}. These resources are valuable for constructing prior knowledge spaces. Static database relations alone, however, do not solve patient-specific syndrome differentiation or prescription selection. We therefore treat these resources as priors that must be activated and weighted by patient evidence.

TCM-specific benchmark datasets provide a complementary perspective. TCM-SD supports syndrome differentiation from clinical narratives \cite{ren2022tcmsd}, while TCM-BEST4SDT evaluates syndrome differentiation and treatment reasoning with prescription information \cite{li2025tcmbest4sdt}. These datasets move TCM decision support from database lookup toward supervised and benchmarked modelling. However, most benchmark settings still evaluate label or text prediction. They do not explicitly model how symptoms, syndromes, treatment principles, herbs, and doses are connected through patient-conditioned reasoning paths. Our framework addresses this gap by making the intermediate H1 and H2 paths part of both prediction and audit.

\subsection{Graph and Hypergraph Learning}

Graph neural networks and relational graph encoders model structured dependencies through pairwise edges \cite{kipf2017gcn,hamilton2017graphsage,schlichtkrull2018rgcn}. Hypergraph learning extends this idea to higher-order relations \cite{zhou2006hypergraph}, and hypergraph neural networks have been used to model group-wise dependencies \cite{feng2019hgnn}. Dynamic hypergraph models further adapt graph structure or hyperedge incidence weights to input instances \cite{jiang2019dhgnn}. TCM diagnosis and prescription are naturally higher-order. Symptoms form patterns, therapeutic principles connect to herb roles, and formulas involve coordinated herb groups. We therefore use hypergraphs to represent diagnostic and prescription paths.

Prior-guided representation and generation have also been studied in other structured domains. Examples include transformer-based brain-tumor segmentation with prior knowledge and text-to-3D generation guided by prior knowledge \cite{li2023brainprior,nie2024t2td}. Our work differs by using patient-conditioned weights over two coupled hypergraphs. In this setting, prior knowledge is not only injected into representation learning, but is also exposed as auditable diagnostic and prescription paths.

\subsection{Interpretability and Consistency in Medical AI}

Interpretability methods such as LIME and SHAP explain model predictions through local feature attribution \cite{ribeiro2016lime,lundberg2017shap}. In medical AI, explanation quality also depends on whether the explanation follows domain-relevant evidence paths. For TCM prescription support, a useful explanation should connect symptoms to syndromes, syndromes to therapeutic principles, and therapeutic principles to herbs. This requirement motivates path-level consistency: the prediction should be compatible with an inspectable route through prior knowledge. Our framework operationalizes this idea with H1 and H2 reasoning paths, path-constrained decoding, and ablation controls that test whether these paths contribute beyond patient text features alone.

\section{Method}

\subsection{Overview}

We design a two-stage decision-support framework. Stage 1 performs syndrome differentiation from clinical narratives with H1, a patient-conditioned diagnostic hypergraph. Stage 2 predicts prescription herb sets and matched-herb doses with H2, a patient-conditioned prescription hypergraph with dose priors. Language models provide patient text representations. H1 and H2 provide structured paths whose weights are adapted to each patient. The system is intended for computational modelling and clinician-reviewable decision support, not autonomous prescription.

\begin{figure*}[t]
  \centering
  \includegraphics[width=0.96\textwidth]{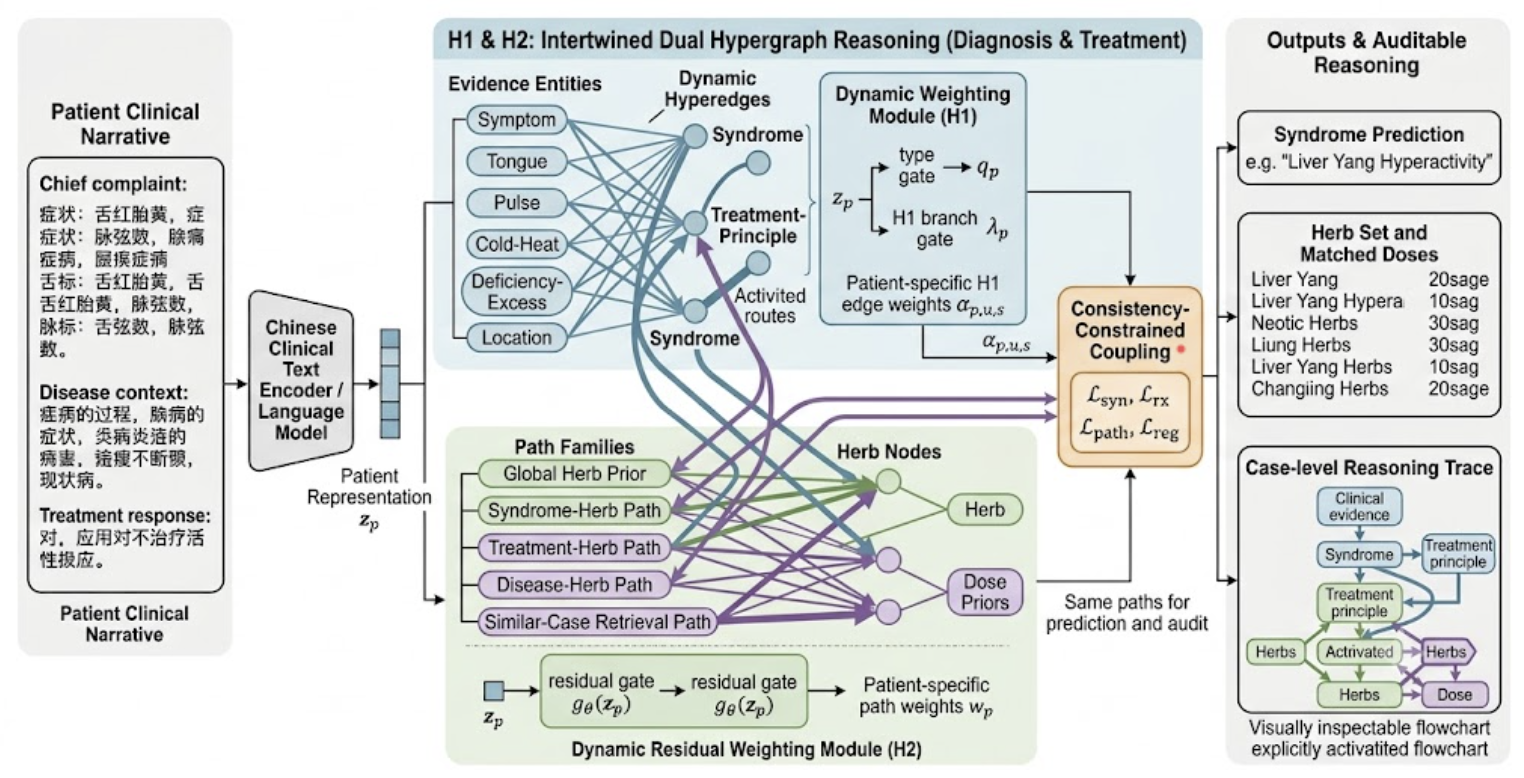}
  \caption{Overview of the proposed patient-conditioned dual hypergraph framework. A patient narrative is encoded into a patient representation that dynamically modulates H1 symptom--syndrome--treatment hyperedge incidences and H2 treatment--herb--dose path weights. H1 supports syndrome and treatment-principle reasoning, while H2 performs path-constrained herb and dose prediction. The same activated paths are used for scoring, prescription assembly, and case-level audit, enabling patient-adaptive and inspectable TCM prescription support.}
  \label{fig:method}
\end{figure*}

\subsection{Problem Formulation}

For patient \(p\), let \(x_p\) denote the clinical narrative, \(s_p\) the syndrome, \(t_p\) the therapeutic principle, and \(Y_p=\{(h_i,d_i)\}_{i=1}^{n_p}\) the reference prescription. The framework models the structured transition
\begin{equation}
x_p \rightarrow z_p \rightarrow (s_p,t_p) \rightarrow Y_p \rightarrow R_p ,
\end{equation}
where \(z_p\) is the patient representation and \(R_p\) is the inspectable reasoning trace assembled from selected H1 and H2 evidence paths. Stage 1 estimates \(P(s\mid x_p,\mathcal{H}_1)\). Stage 2 estimates herb scores and doses conditioned on \(x_p\), syndrome/treatment evidence, and \(\mathcal{H}_2\).

\subsection{Patient Text Representation}

A Chinese-capable language model or pretrained encoder maps the narrative into a representation
\begin{equation}
\mathbf{z}_p = E_\psi(x_p).
\end{equation}
In the completed experiments, BERT-base Chinese, MacBERT, RoBERTa-wwm-ext, and a Medical/TCM-domain BERT are used as neural text encoders. Qwen2.5 is evaluated in zero-shot, few-shot, candidate-constrained, retrieval-augmented, and prescription-generation settings. These language-model baselines are used for comparison, while the proposed model uses language representations as structured components rather than as autonomous prescribers.

\subsection{First Hypergraph for Syndrome Reasoning}

The first hypergraph is
\begin{equation}
\mathcal{H}_1=(\mathcal{V}_1,\mathcal{E}_1),
\end{equation}
where the static H1 node set includes symptom, tongue, pulse, cold--heat, deficiency--excess, location, and syndrome nodes. During patient-specific inference, this static structure induces \(\mathcal{H}_1^{(p)}=(\mathcal{V}_1\cup\{v_p\}\cup\mathcal{T}_p,\mathcal{E}_1^{(p)})\), where \(v_p\) is a patient-instance node and \(\mathcal{T}_p\) contains treatment-principle bridge nodes when they are available. H1 is not a pairwise entity--syndrome graph. For a patient \(p\) and candidate syndrome \(s\), an instantiated diagnostic hyperedge is
\begin{equation}
e_{p,s}^{H1}
=
\{v_p,s\}\cup \mathcal{U}_{p,s}\cup \mathcal{T}_{s},
\end{equation}
where \(v_p\) is the patient node, \(\mathcal{U}_{p,s}\) is the set of patient evidence entities connected to syndrome \(s\), and \(\mathcal{T}_{s}\) denotes treatment-principle bridge nodes associated with \(s\) when they are observed or available from the knowledge source. Thus, an H1 hyperedge contains at least the patient, multiple clinical evidence entities, and a syndrome node; in the treatment-annotated setting it also connects to treatment-principle nodes. The implemented incidence weights are stored in \(\mathbf{A}_1\), where \(A_{1,u,s}\) denotes the incidence strength between evidence entity \(u\) and syndrome \(s\) inside the larger hyperedge \(e_{p,s}^{H1}\). For patient \(p\), extracted entity features form a vector \(\mathbf{e}_p\). The patient representation generates type-level and branch-level gates:
\begin{equation}
\mathbf{q}_p=\sigma(f_{\theta}^{\mathrm{type}}(\mathbf{z}_p)),\quad
\lambda_p=\sigma(f_{\theta}^{\mathrm{H1}}(\mathbf{z}_p)),
\end{equation}
where \(\mathbf{q}_p\) weights entity groups such as symptom, tongue, pulse, cold--heat, deficiency--excess, and location. If \(\tau(u)\) is the entity type of \(u\), the patient-specific H1 incidence weight is
\begin{equation}
\alpha_{p,u,s}
=
e_{p,u}\,q_{p,\tau(u)}\,\mathrm{softplus}(A_{1,u,s}).
\end{equation}
The H1 evidence logit for syndrome \(s\) is the sum of activated incidences inside the patient-specific hyperedge,
\begin{equation}
r_{p,s}^{H1}=\sum_{u\in \mathcal{U}_{p,s}}\alpha_{p,u,s},
\end{equation}
and all syndrome labels are scored jointly by
\begin{equation}
\operatorname{Score}_{H1}(p,s)=\operatorname{logit}_{\theta}(x_p,s)+\lambda_p r_{p,s}^{H1}.
\end{equation}
Thus, the constructed H1 topology is shared across patients, but the instantiated H1 hyperedge and its incidence weights are individualized by the patient's observed entities and patient-specific gates. The matrix \(\mathbf{A}_1\), the type gate \(f_{\theta}^{\mathrm{type}}\), and the branch gate \(f_{\theta}^{\mathrm{H1}}\) are learned jointly through the objective in Section~III-F.

\subsection{Second Hypergraph for Prescription Reasoning}

The second hypergraph is
\begin{equation}
\mathcal{H}_2=(\mathcal{V}_2,\mathcal{E}_2),
\end{equation}
where the static H2 node set includes syndrome nodes, treatment-principle nodes, and herb nodes, with herb-dose priors stored as attributes for dose evaluation. During patient-specific inference, the static structure induces \(\mathcal{H}_2^{(p)}\) by adding \(v_p\), the disease context \(c_p\), and retrieved training cases as transient conditioning nodes. H2 contains prescription hyperedges that connect a patient context to candidate herbs through multiple clinical priors. For patient \(p\) and candidate herb \(h\), the instantiated H2 support hyperedges are
\begin{equation}
\begin{aligned}
e_{p,h}^{\mathrm{syn}} &= \{v_p,s_p,h,d_h\},\\
e_{p,h}^{\mathrm{trt}} &= \{v_p,t_p,h,d_h\},\\
e_{p,h}^{\mathrm{dis}} &= \{v_p,c_p,h,d_h\},\\
e_{p,h}^{\mathrm{ret}} &= \{v_p,r_{p,1},\ldots,r_{p,K},h,d_h\},
\end{aligned}
\end{equation}
where \(s_p\) is the syndrome, \(t_p\) is the treatment principle, \(c_p\) is the disease context when available, \(r_{p,1},\ldots,r_{p,K}\) are retrieved similar training cases, and \(d_h\) is the herb dose prior. The global herb prior is represented as a herb--dose prior incidence and is combined with these patient-conditioned hyperedges during scoring. H2 path features include a global herb prior, syndrome--herb prior, treatment--herb prior, disease--herb prior, and similar-case retrieval path.

For each patient--herb pair \((p,h)\), the path feature vector is
\begin{equation}
\boldsymbol{\phi}_{p,h}
=
[\phi^{\mathrm{global}}_{h},
\phi^{\mathrm{syn}}_{p,h},
\phi^{\mathrm{trt}}_{p,h},
\phi^{\mathrm{dis}}_{p,h},
\phi^{\mathrm{ret}}_{p,h}].
\end{equation}
The learned fixed H2 model uses a global nonnegative path-weight vector \(\mathbf{w}_0\). The proposed patient-conditioned residual H2 model adapts this vector with a bounded residual gate:
\begin{equation}
\mathbf{w}_p = \mathbf{w}_0 \odot \left(1 + 0.5\tanh(g_\theta(\mathbf{z}_p))\right),
\end{equation}
where \(w_{p,k}\) is the patient-specific weight for the \(k\)-th H2 path family. We use
\begin{equation}
\gamma_{p,k}=w_{p,k}\phi_{p,h}^{k}
\end{equation}
as the activated support carried by path family \(k\) for herb \(h\), and score herbs as
\begin{equation}
\operatorname{Score}_{H2}(p,h)=\mathbf{w}_p^\top \boldsymbol{\phi}_{p,h}.
\end{equation}
The bounded residual design keeps the hypergraph prior as an anchor while allowing patient-specific adjustment. In both H1 and H2, the method learns individualized hypergraph weights without replacing the constructed knowledge structure with unconstrained attention.

\subsection{Path-Consistent Learning and Inference}

The framework uses consistency in two senses. First, syndrome and herb scores are computed through H1 or H2 evidence paths with patient-specific weights. Second, explanation traces are generated from the same paths used for scoring, rather than from a separate post-hoc prompt. For syndrome label \(s\), define the strongest H1 path support as
\begin{equation}
\rho_{p,s}^{H1}=\max_{\pi\in\Pi_1(s)} \prod_{u\in\pi\cap\mathcal{U}_{p,s}} \alpha_{p,u,s},
\end{equation}
where \(\Pi_1(s)\) is the set of H1 evidence paths inside the instantiated hyperedge for \(s\). For herb \(h\), define H2 support as
\begin{equation}
\rho_{p,h}^{H2}=\max_{\pi\in\Pi_2(h)} \prod_{k\in\pi} \gamma_{p,k}.
\end{equation}
These supports are used in a four-term training objective. The syndrome classification term is
\begin{equation}
\mathcal{L}_{\mathrm{syn}}
=
-\sum_p \log P_\theta(s_p^\ast\mid x_p,\mathcal{H}_1),
\end{equation}
where \(s_p^\ast\) is the reference syndrome. The prescription supervision term is a single objective that combines herb-set prediction and matched-dose fitting:
\begin{equation}
\mathcal{L}_{\mathrm{rx}}
=
\mathcal{B}_{\mathrm{sel}}+\beta_d\mathcal{D}_{\mathrm{dose}},
\end{equation}
where
\begin{equation}
\begin{aligned}
\mathcal{B}_{\mathrm{sel}}
=
-\sum_p\sum_h
\big[
&y_{p,h}\log\sigma(o_{p,h})\\
&+(1-y_{p,h})\log(1-\sigma(o_{p,h}))
\big],
\end{aligned}
\end{equation}
and
\begin{equation}
\mathcal{D}_{\mathrm{dose}}
=
\sum_p\sum_{h\in \hat{Y}_p^H\cap Y_p^H}
\left|\hat{d}_{p,h}-d_{p,h}\right|.
\end{equation}
Here \(o_{p,h}=\operatorname{Score}_{H2}(p,h)\). Thus, \(\mathcal{L}_{\mathrm{rx}}\) learns herb selection and calibrates the predicted dose for herbs that match the reference prescription.
The path-consistency term constrains the correct syndrome and reference herbs to be supported by activated H1 and H2 paths:
\begin{equation}
\mathcal{L}_{\mathrm{path}}
=
-\sum_p \log(\rho_{p,s_p^\ast}^{H1}+\epsilon)
-\sum_p \sum_{h\in Y_p^H}
\log(\rho_{p,h}^{H2}+\epsilon).
\end{equation}
where \(\epsilon\) is a small constant for numerical stability. This term rewards activated routes that explain the reference syndrome and herbs, and it provides supervision for the dynamic H1 incidence function and H2 path-weight function.
We use a regularization term to anchor dynamic weights to the constructed priors and keep patient-specific gates stable:
\begin{equation}
\begin{aligned}
\mathcal{L}_{\mathrm{reg}}
=
&\|\mathbf{A}_1-\mathbf{A}_1^{0}\|_F^2
+\sum_p\|\mathbf{w}_p-\mathbf{w}_0\|_2^2\\
&+\sum_p \|\mathbf{q}_p\|_1
+\sum_p\|g_\theta(\mathbf{z}_p)\|_2^2.
\end{aligned}
\end{equation}
where \(\mathbf{A}_1^0\) is the initialized H1 incidence matrix. The first two parts keep the learned H1 incidence matrix and the patient-specific H2 weights close to the constructed priors. The last two parts keep the patient gates sparse and numerically stable. This regularization helps the individualized weights remain interpretable rather than becoming unconstrained attention.
The complete objective is
\begin{equation}
\mathcal{L}
=
\mathcal{L}_{\mathrm{syn}}
+\mathcal{L}_{\mathrm{rx}}
+\beta_p\mathcal{L}_{\mathrm{path}}
+\beta_r\mathcal{L}_{\mathrm{reg}}.
\end{equation}
The four terms have distinct roles. \(\mathcal{L}_{\mathrm{syn}}\) learns syndrome prediction through H1. \(\mathcal{L}_{\mathrm{rx}}\) learns prescription composition and dose fitting through H2. \(\mathcal{L}_{\mathrm{path}}\) encourages the dynamic edge and path-weight functions to assign mass to clinically coherent H1 and H2 routes. \(\mathcal{L}_{\mathrm{reg}}\) keeps patient-specific modulation anchored to the constructed hypergraphs. At inference time, \(R_p\) is formed by selecting the highest-support H1 and H2 paths under \(\rho_{p,s}^{H1}\) and \(\rho_{p,h}^{H2}\). Thus, the same dynamic weights drive scoring, constrained prescription assembly, and explanation.

\section{Experiments}

TCM-SD contains clinical records with chief complaints, disease descriptions, four-diagnostic information, and normalized syndrome labels \cite{ren2022tcmsd}. We use the public train/dev split with 43,180 training records, 5,486 development records, and 148 syndrome labels. Disease shortcut fields (\texttt{lcd\_id} and \texttt{lcd\_name}) are excluded from neural classifier inputs.

TCM-BEST4SDT contains syndrome differentiation and treatment reasoning cases for evaluating large language models \cite{li2025tcmbest4sdt}. We use 300 TCM-SDT prescription cases. Each case includes clinical text, syndrome type, etiology/pathogenesis fields, therapeutic principles, and herb composition with doses. Five-fold cross-validation is used for prescription experiments.

SymMap and ETCM are used as prior knowledge sources rather than patient-level supervised datasets. They provide symptoms, syndromes, diseases, formulas, herbs, and formula-level relations for constructing H1 and H2 evidence paths. Table~\ref{tab:data} summarizes the experimental sources. Table~\ref{tab:hypergraph-stats} reports the constructed H1 and H2 hypergraph sizes.

\begin{table}[t]
\centering
\caption{Datasets and Knowledge Sources}
\label{tab:data}
\begin{tabular}{p{0.22\columnwidth}p{0.22\columnwidth}p{0.43\columnwidth}}
\toprule
Source & Scale & Use in this study \\
\midrule
TCM-SD & 43,180 train; 5,486 dev; 148 syndrome labels & Stage 1 syndrome differentiation from clinical narratives. \\
TCM-BEST4SDT & 300 prescription cases & Stage 2 five-fold herb-set and matched-dose evaluation. \\
SymMap & External prior database & Symptom, syndrome, disease, and herb prior paths. \\
ETCM & External prior database & Formula and herb prior paths. \\
\bottomrule
\end{tabular}
\end{table}

\begin{table*}[t]
\centering
\scriptsize
\caption{Constructed H1 and H2 Hypergraph Summary}
\label{tab:hypergraph-stats}
\resizebox{\textwidth}{!}{%
\begin{tabular}{p{0.07\textwidth}p{0.18\textwidth}p{0.19\textwidth}p{0.25\textwidth}p{0.19\textwidth}}
\toprule
Graph & Construction basis & Node inventory & Hyperedge template and incidence statistics & Example instantiated hyperedges \\
\midrule
H1 & TCM-SD clinical text, 148 syndrome labels, and symptom/tongue/pulse/cold--heat/deficiency--excess/location entities & Static: 148 syndrome nodes and 151 evidence-entity nodes. Dynamic: patient node and optional treatment bridge nodes. & Diagnostic hyperedge: \(\{v_p,s,\mathcal{U}_{p,s},\mathcal{T}_s\}\). Static incidence table: 9,718 evidence--syndrome incidences; mean 65.7 per syndrome. & \(\{p,\) wiry pulse, white coating, upper-abdominal evidence, liver-qi stagnation\(\}\); \(\{p,\) fever, red tongue, rash, heat-toxin exuberance\(\}\) \\
H2 & TCM-BEST4SDT prescriptions, syndrome labels, treatment principles, disease context, herbs, and dose records & Static: 267 treatment-principle nodes, 228 syndrome nodes, and 529 herb nodes. Dynamic: patient, disease-context, and retrieved-case nodes. & Prescription hyperedges: \(\{v_p,s_p,h,d_h\}\), \(\{v_p,t_p,h,d_h\}\), \(\{v_p,c_p,h,d_h\}\), and retrieval \(\{v_p,r_{1:K},h,d_h\}\). Static incidence table: 5,131 prior incidences. & \(\{p,\) soothing liver/regulating qi, Chaihu, dose prior\(\}\); \(\{p,\) clearing heat/detoxifying, Huangqin, dose prior\(\}\) \\
\bottomrule
\end{tabular}
}
\end{table*}

Stage 1 is evaluated with accuracy, macro-F1, and top-3 accuracy. Stage 2 is evaluated with herb precision, recall, F1, prescription Jaccard similarity, and mean absolute error (MAE) on matched herbs. For each Step 2 test case, the number of predicted herbs is set to the reference herb count so that herb precision, recall, and F1 measure herb-selection quality under matched prescription size. Prescription Jaccard is
\begin{equation}
\mathrm{Jaccard}(\hat{Y}_p^H,Y_p^H)
=
\frac{|\hat{Y}_p^H\cap Y_p^H|}
{|\hat{Y}_p^H\cup Y_p^H|},
\end{equation}
where \(\hat{Y}_p^H\) and \(Y_p^H\) are predicted and reference herb sets.

\subsection{Main Results}

We compare the proposed dual-hypergraph framework with traditional, neural, retrieval, knowledge-guided, hypergraph, and language-model baselines. The evaluation is organized at three levels: Stage 1 syndrome differentiation, Stage 2 prescription prediction, and the connected Step 1-to-Step 2 pipeline. The final baseline set includes frequency and TF-IDF methods, CNN and recurrent neural models, pretrained Chinese encoders, entity-enhanced and retrieval-enhanced encoders, knowledge graph and static hypergraph baselines, Qwen2.5 variants, fixed-prior H2 models, and patient-conditioned H1/H2 models.

\subsubsection{Stage 1 syndrome differentiation}

Table~\ref{tab:step1} compares Stage 1 syndrome differentiation on TCM-SD. Traditional and shallow neural methods performed poorly: majority prediction, TF-IDF+SVM, and TextCNN all remained below 0.315 accuracy, while BiLSTM-Attention reached 0.4851. Pretrained Chinese encoders provided the main performance jump. BERT-base Chinese reached 0.7973 accuracy, 0.3047 macro-F1, and 0.9061 top-3 accuracy. MacBERT alone reached similar accuracy but lower macro-F1. Entity-enhanced and retrieval-enhanced variants improved over plain encoders, showing that explicit clinical evidence and similar cases are useful for syndrome differentiation.

The best Stage 1 result was obtained by MacBERT with dynamic H1 weights, reaching 0.8297 accuracy, 0.3288 macro-F1, and 0.9167 top-3 accuracy. Compared with MacBERT alone, dynamic H1 improved accuracy by 3.82 percentage points and macro-F1 by 5.45 points. Compared with MacBERT plus fixed H1 weights, dynamic H1 improved accuracy by 1.79 points and macro-F1 by 1.83 points. Qwen2.5 improved substantially when constrained by candidate syndrome labels and retrieval, but the best RAG-constrained Qwen2.5 result (0.7819 accuracy and 0.3037 macro-F1) remained below the best dynamic H1 model. This supports the role of H1 as a patient-conditioned evidence structure rather than a static label prior.

\begin{table*}[t]
\centering
\scriptsize
\caption{Stage 1 Syndrome Differentiation on TCM-SD}
\label{tab:step1}
\resizebox{\textwidth}{!}{%
\begin{tabular}{llccc}
\toprule
Method & Category & Accuracy & Macro-F1 & Top-3 Acc. \\
\midrule
Majority / frequency label & Traditional & 0.2834 & 0.1975 & 0.4833 \\
TF-IDF + SVM & Traditional & 0.2793 & 0.2007 & 0.4815 \\
TextCNN & Neural & 0.3147 & 0.2016 & 0.5012 \\
BiLSTM-Attention & Neural & 0.4851 & 0.2538 & 0.6603 \\
BERT-base Chinese & Text encoder & 0.7973 & 0.3047 & 0.9061 \\
MacBERT & Text encoder & 0.7915 & 0.2743 & 0.9021 \\
RoBERTa-wwm-ext & Text encoder & 0.7884 & 0.2876 & 0.9000 \\
Medical / TCM-domain BERT & Domain encoder & 0.8072 & 0.2905 & 0.9088 \\
BERT + extracted entities & Entity-enhanced & 0.8095 & 0.3118 & 0.9012 \\
MacBERT + extracted entities & Entity-enhanced & 0.8176 & 0.3169 & 0.9001 \\
BERT + retrieved similar cases & Retrieval-enhanced & 0.8138 & 0.3145 & 0.9026 \\
MacBERT + retrieved similar cases & Retrieval-enhanced & 0.8114 & 0.3117 & 0.9052 \\
Qwen2.5 zero-shot & LLM & 0.5842 & 0.2386 & 0.7315 \\
Qwen2.5 few-shot & LLM & 0.6418 & 0.2569 & 0.7863 \\
Qwen2.5 + candidate syndrome list & Constrained LLM & 0.7426 & 0.2864 & 0.8755 \\
Qwen2.5 RAG + candidate syndrome list & RAG LLM & 0.7819 & 0.3037 & 0.9016 \\
H1 evidence only & Knowledge-only & 0.3270 & 0.2172 & 0.5361 \\
BERT + fixed H1 weights & Ours / ablation & 0.8002 & 0.3086 & 0.9105 \\
BERT + dynamic H1 weights & Ours & 0.7993 & 0.3020 & 0.9063 \\
MacBERT + fixed H1 weights & Ours / ablation & 0.8118 & 0.3105 & 0.9130 \\
MacBERT + dynamic H1 weights & Ours & \textbf{0.8297} & \textbf{0.3288} & \textbf{0.9167} \\
\bottomrule
\end{tabular}
}
\end{table*}

Overall, Table~\ref{tab:step1} shows that knowledge-only H1 is insufficient by itself, but dynamic H1 improves the strongest MacBERT encoder. The result suggests that H1 is most useful when it modulates a strong patient text representation rather than replacing the encoder.

\subsubsection{Stage 2 prescription prediction}

Table~\ref{tab:step2} reports five-fold prescription prediction on 300 TCM-BEST4SDT cases. Simple frequency baselines were weak, ranging from 0.1808 to 0.2348 Herb-F1. Retrieval was much stronger in this low-resource setting: BM25, dense retrieval, and similar-case retrieval reached 0.2742, 0.2897, and 0.3059 Herb-F1, respectively. Neural multi-label classifiers improved when syndrome and treatment information were provided, with the BERT syndrome+treatment classifier reaching 0.3018 and the label-wise attention classifier reaching 0.3036.

Knowledge-guided models were competitive. The R-GCN baseline reached 0.2956 Herb-F1, and the static HGNN reached 0.3071. The learned fixed H2 model reached 0.3085, while the proposed dynamic residual H2 reached 0.3090 in the seed-42 run and 0.3111 as the three-seed mean. This is the highest Herb-F1 in the final comparison. Strong constrained language-model baselines were close but slightly lower: Qwen2.5 with RAG and candidate-herb reranking reached 0.3078 Herb-F1. Direct Qwen2.5 generation remained weaker at 0.2577, indicating that unconstrained generation is not well aligned with exact reference herb sets.

\begin{table*}[t]
\centering
\scriptsize
\caption{Stage 2 Prescription Prediction on TCM-BEST4SDT}
\label{tab:step2}
\resizebox{\textwidth}{!}{%
\begin{tabular}{llccc}
\toprule
Method & Category & Herb-F1 & Jaccard & Dose MAE \\
\midrule
Herb frequency & Traditional & 0.1808 & 0.1056 & \textbf{2.9010} \\
Syndrome-specific herb frequency & Traditional & 0.2146 & 0.1327 & 3.0418 \\
Treatment-specific herb frequency & Traditional & 0.2219 & 0.1398 & 3.0875 \\
Syndrome + treatment herb frequency & Traditional & 0.2348 & 0.1486 & 3.1652 \\
BM25 retrieval & Retrieval & 0.2742 & 0.1861 & 3.4317 \\
Dense retrieval & Retrieval & 0.2897 & 0.1984 & 3.5028 \\
Similar-case retrieval & Retrieval & 0.3059 & \textbf{0.2119} & 3.5589 \\
BERT multi-label herb classifier & Neural & 0.2815 & 0.1912 & 3.6031 \\
BERT + syndrome multi-label classifier & Neural & 0.2928 & 0.1997 & 3.6684 \\
BERT + treatment multi-label classifier & Neural & 0.2899 & 0.1979 & 3.6522 \\
BERT + syndrome + treatment multi-label classifier & Neural & 0.3018 & 0.2064 & 3.7245 \\
Label-wise attention classifier & Neural & 0.3036 & 0.2075 & 3.7481 \\
R-GCN / knowledge graph baseline & Knowledge graph & 0.2956 & 0.2018 & 3.6174 \\
Static HGNN / hypergraph encoder & Hypergraph & 0.3071 & 0.2081 & 3.7716 \\
Fixed H2 prior & Knowledge prior & 0.2422 & 0.1551 & 3.2139 \\
Learned fixed H2 & Ours / ablation & 0.3085 & 0.2088 & 3.7933 \\
Dynamic residual H2 & Ours & 0.3090 & 0.2096 & 3.8229 \\
Dynamic residual H2, 3-seed mean & Ours & \textbf{0.3111} & 0.2107 & 3.8064 \\
Qwen2.5 direct generation & LLM & 0.2577 & 0.1704 & 4.1268 \\
Qwen2.5 few-shot generation & LLM & 0.2738 & 0.1821 & 4.0025 \\
Qwen2.5 RAG generation & RAG LLM & 0.2869 & 0.1916 & 3.9188 \\
Qwen2.5 + candidate herb list & Constrained LLM & 0.2968 & 0.2007 & 3.8611 \\
Qwen2.5 candidate-herb reranker & Constrained LLM & 0.3027 & 0.2053 & 3.8124 \\
Qwen2.5 RAG + candidate-herb reranker & Strong LLM baseline & 0.3078 & 0.2091 & 3.7762 \\
\bottomrule
\end{tabular}
}
\end{table*}

Overall, Table~\ref{tab:step2} shows that similar-case retrieval and constrained language-model reranking are strong baselines, but dynamic residual H2 gives the best Herb-F1. The small margin also shows that the prescription task is difficult and that H2 should be interpreted as a structured improvement over strong retrieval and constrained-generation baselines, not as a large absolute-performance jump.

\begin{figure*}[t]
  \centering
  \includegraphics[width=0.96\textwidth,height=0.55\textheight,keepaspectratio]{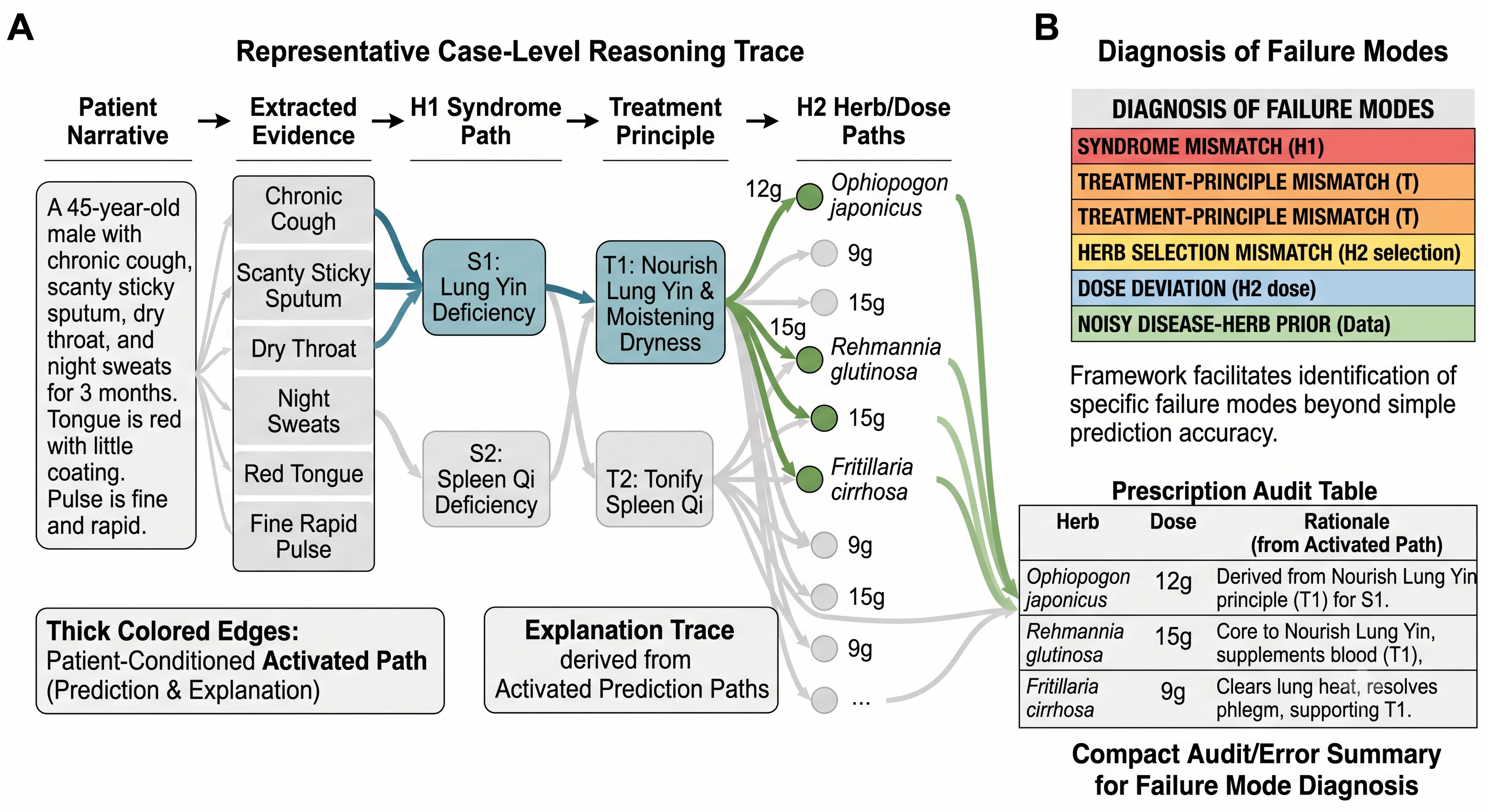}
  \caption{Representative case-level audit trace for prescription prediction. The audit view links patient evidence, H1 syndrome reasoning, treatment-principle information, H2 herb support paths, and predicted-versus-reference herb sets. The example illustrates how matched, missed, and extra herbs can be inspected together with their supporting reasoning paths; it does not constitute evidence of clinical efficacy.}
  \label{fig:case}
\end{figure*}

Figure~\ref{fig:case} shows a representative held-out case-level audit trace. This example is not clinical validation. It illustrates how predicted syndrome information, treatment principles, herb-level errors, and H2 path support can be inspected in one view.

\subsubsection{Connected Step 1-to-Step 2 evaluation}

Table~\ref{tab:e2e} reports the connected Step 1-to-Step 2 evaluation. The oracle setting with gold syndrome and gold treatment reached 0.3101 Herb-F1. Replacing the gold syndrome with H1-predicted syndrome while retaining gold treatment reached 0.3043, showing that high-quality syndrome prediction preserves most of the oracle prescription performance. Removing gold treatment reduced Herb-F1 to 0.2980, and mapping the predicted syndrome to the most frequent training-fold treatment reduced it further to 0.2550. Thus, treatment-principle information remains important, and naive syndrome-to-treatment mapping is still a weak bridge.

Across full pipelines, the proposed Dynamic H1 + Dynamic H2 setting achieved 0.3074 Herb-F1, 0.2084 Jaccard, and 3.8268 Dose MAE. This result is close to the oracle H2 setting and higher than BERT+dynamic H2, MacBERT+retrieval, Qwen2.5-only generation, and RAG diagnosis plus RAG prescription. The result supports the two-stage design: dynamic H1 improves syndrome prediction, and dynamic H2 uses the resulting structured context for prescription inference.

\begin{table*}[t]
\centering
\caption{End-to-End Step 1-to-Step 2 Evaluation. Syn. Top-1/Top-3 are computed on the TCM-BEST4SDT connected-evaluation subset after syndrome normalization.}
\label{tab:e2e}
\scriptsize
\resizebox{\textwidth}{!}{%
\begin{tabular}{lllccccc}
\toprule
Setting & Step 1 Source & Step 2 Source & Herb-F1 & Jaccard & Dose MAE & Syn. Top-1 & Syn. Top-3 \\
\midrule
Gold syndrome + gold treatment + H2 & Gold & Gold treatment + H2 & \textbf{0.3101} & \textbf{0.2102} & 3.8183 & -- & -- \\
Predicted syndrome + gold treatment + H2 & H1 predicted & Gold treatment + H2 & 0.3043 & 0.2049 & 3.8391 & 0.8712 & 0.9033 \\
Predicted syndrome + blank treatment + H2 & H1 predicted & Blank treatment + H2 & 0.2980 & 0.1998 & 3.9229 & 0.8712 & 0.9033 \\
Predicted syndrome + mapped treatment + H2 & H1 predicted & Frequent mapped treatment + H2 & 0.2550 & 0.1705 & 3.9553 & 0.8712 & 0.9033 \\
BERT Step 1 + frequency Step 2 & BERT & Frequency & 0.2136 & 0.1332 & \textbf{3.1284} & 0.7973 & 0.9061 \\
BERT Step 1 + retrieval Step 2 & BERT & Retrieval & 0.2817 & 0.1904 & 3.5267 & 0.7973 & 0.9061 \\
BERT Step 1 + learned fixed H2 & BERT & Learned fixed H2 & 0.2968 & 0.2009 & 3.8175 & 0.7973 & 0.9061 \\
BERT Step 1 + dynamic H2 & BERT & Dynamic H2 & 0.2995 & 0.2026 & 3.8438 & 0.7973 & 0.9061 \\
MacBERT Step 1 + retrieval Step 2 & MacBERT & Retrieval & 0.2869 & 0.1947 & 3.5112 & 0.7915 & 0.9021 \\
MacBERT Step 1 + dynamic H2 & MacBERT & Dynamic H2 & 0.3058 & 0.2067 & 3.8316 & 0.8297 & 0.9167 \\
Qwen2.5 Step 1 + Qwen2.5 Step 2 & LLM & LLM generation & 0.2446 & 0.1588 & 4.2165 & 0.5842 & 0.7315 \\
Qwen2.5 Step 1 + retrieval-constrained Qwen2.5 Step 2 & LLM & RAG / constrained LLM & 0.2735 & 0.1819 & 4.0287 & 0.5842 & 0.7315 \\
Qwen2.5 Step 1 + H2 & LLM & Dynamic H2 & 0.2862 & 0.1915 & 3.9674 & 0.5842 & 0.7315 \\
H1 Step 1 + Qwen2.5 candidate reranker & H1 predicted & LLM reranker & 0.3006 & 0.2038 & 3.8492 & 0.8712 & 0.9033 \\
RAG diagnosis + RAG prescription & Retrieval / RAG & Retrieval / RAG & 0.2927 & 0.1972 & 3.9145 & 0.7819 & 0.9016 \\
Full proposed pipeline & Dynamic H1 & Dynamic H2 & 0.3074 & 0.2084 & 3.8268 & 0.8297 & \textbf{0.9167} \\
\bottomrule
\end{tabular}
}
\end{table*}

The connected evaluation shows that most prescription performance is retained when the predicted syndrome is combined with gold treatment. In contrast, a simple predicted-syndrome-to-treatment mapping is not reliable. This finding motivates the ablation analysis below. The remaining bottleneck is not only syndrome prediction, but also the semantic bridge from syndrome labels to treatment principles and dose-safe herb selection.

\subsection{Ablation Studies}

We next analyze which components are responsible for the observed behavior. The ablation studies are organized in the same three levels as the main comparison: Stage 1, Stage 2, and the connected pipeline.

\subsubsection{Stage 1 H1 ablation}

Table~\ref{tab:step1} contains the Stage 1 ablation evidence. H1 evidence alone reached only 0.3270 accuracy and 0.2172 macro-F1, far below pretrained text encoders. This confirms that H1 is not intended to replace clinical text representation. Its role is to provide a structured evidence space around a strong encoder. With BERT-base Chinese, fixed H1 weights slightly improved accuracy and top-3 accuracy over BERT alone, while dynamic H1 did not improve macro-F1. With MacBERT, however, the pattern changed: fixed H1 improved accuracy from 0.7915 to 0.8118, and dynamic H1 further improved accuracy to 0.8297 and macro-F1 to 0.3288. This suggests that patient-conditioned H1 weighting is beneficial when the underlying encoder provides sufficiently stable clinical representations.

\begin{figure}[t]
  \centering
  \includegraphics[width=\columnwidth]{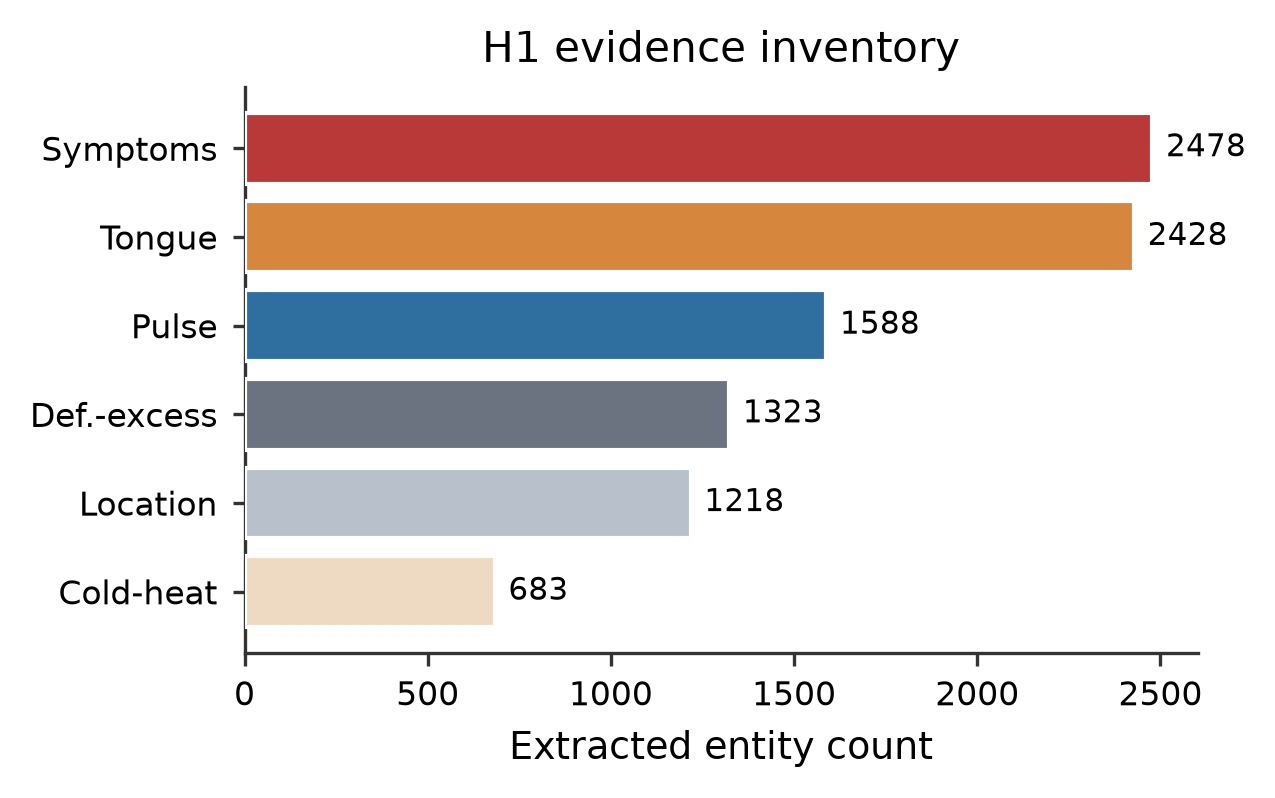}
  \caption{H1 entity inventory used for evidence-path construction.}
  \label{fig:step1-inventory}
\end{figure}

Figure~\ref{fig:step1-inventory} reports the evidence inventory underlying H1 path weighting. Symptoms and tongue entities dominate the extracted evidence space, followed by pulse and deficiency--excess features. This inventory matters because the audit trace depends on observable evidence paths rather than latent text scores alone.

\subsubsection{Stage 2 H2 ablation}

The improvement from learned fixed H2 to dynamic residual H2 is small, but it is directionally stable across three seeds (Table~\ref{tab:seed}). Mean Herb-F1 is 0.3111 for dynamic residual H2, 0.3095 for learned fixed H2, 0.2980 for retrieval, 0.2380 for fixed H2 prior, and 0.1784 for frequency. We therefore describe patient-conditioned residual weighting as a consistent but modest improvement.

\begin{table}[t]
\centering
\caption{Seed Stability for Stage 2 Herb-F1}
\label{tab:seed}
\scriptsize
\setlength{\tabcolsep}{3pt}
\resizebox{\columnwidth}{!}{%
\begin{tabular}{lccccc}
\toprule
Method & Mean & Std. & S42 & S43 & S44 \\
\midrule
Frequency & 0.1784 & 0.0032 & 0.1808 & 0.1739 & 0.1806 \\
Retrieval & 0.2980 & 0.0068 & 0.3059 & 0.2894 & 0.2987 \\
Fixed H2 prior & 0.2380 & 0.0036 & 0.2422 & 0.2335 & 0.2383 \\
Learned fixed H2 & 0.3095 & 0.0037 & 0.3085 & 0.3055 & 0.3145 \\
Dynamic residual H2 & \textbf{0.3111} & 0.0031 & 0.3090 & 0.3088 & 0.3155 \\
\bottomrule
\end{tabular}
}
\end{table}

Table~\ref{tab:ablation} reports three-seed mean H2 path ablations using the final all-paths dynamic residual H2 result from Table~\ref{tab:step2} as the reference. Removing retrieval caused the largest drop, reducing Herb-F1 from 0.3111 to 0.2426. This confirms that similar-case evidence is a central path family in the 300-case prescription setting. Removing the treatment--herb path reduced Herb-F1 to 0.3004, and removing the syndrome--herb path reduced it to 0.3033, showing that both structured clinical paths contribute beyond the global prior. Removing the disease--herb path caused only a small decrease to 0.3107, suggesting that disease-level evidence is weaker than treatment and retrieval evidence in this dataset. Removing all H2 path features reduced Herb-F1 to 0.1388; a separate patient-only scorer without H2 paths reached 0.2612 in the same evaluation protocol, still below the full H2 model. These controls indicate that H2 path evidence contributes to herb-set prediction.

\begin{table}[t]
\centering
\caption{H2 Path Ablation for Dynamic Residual H2}
\label{tab:ablation}
\begin{tabular}{lcc}
\toprule
Setting & Herb-F1 & Delta \\
\midrule
All paths & 0.3111 & 0.0000 \\
H2 only & 0.2236 & -0.0875 \\
w/o syndrome--herb & 0.3033 & -0.0078 \\
w/o treatment--herb & 0.3004 & -0.0107 \\
w/o disease--herb & 0.3107 & -0.0004 \\
w/o retrieval path & 0.2426 & -0.0685 \\
No H2 paths & 0.1388 & -0.1723 \\
\bottomrule
\end{tabular}
\end{table}

The ablation pattern shows that retrieval and treatment-level paths carry the strongest incremental signal. The disease--herb path contributes only marginally, which is plausible because disease labels are coarser than treatment principles and may provide less patient-specific prescription guidance. This result illustrates a practical advantage of inspectable path models: they can reveal which knowledge paths are central and which paths mainly provide weak contextual support.

The dose results in Tables~\ref{tab:step2} and \ref{tab:e2e} show that better herb-set prediction does not necessarily imply better dose calibration. Frequency and fixed-prior methods can have lower matched-herb dose MAE, whereas stronger herb-set methods may have larger dose error. This suggests that herb selection and dose estimation should be treated as related but distinct subproblems.

\subsubsection{Connected-pipeline ablation}

The connected-pipeline results in Table~\ref{tab:e2e} show that the proposed Dynamic H1 + Dynamic H2 pipeline retains most of the oracle prescription performance. The full pipeline reached 0.3074 Herb-F1, compared with 0.3101 for the gold syndrome plus gold treatment H2 setting. MacBERT + dynamic H2 reached 0.3058, while BERT + dynamic H2 reached 0.2995. The H1 Step 1 + Qwen2.5 candidate reranker setting reached 0.3006, suggesting that the structured H1 diagnosis is also useful when the prescription generator is an LLM reranker. In contrast, Qwen2.5-only diagnosis and prescription reached 0.2446 Herb-F1, and retrieval-constrained Qwen2.5 reached 0.2735. These results support the connected structured pipeline, while also showing that gold treatment information remains valuable.

\subsection{Case-Level Traces Support Auditability}

Figure~\ref{fig:case} shows a representative held-out case-level audit trace. This example is not clinical validation. It shows how predicted syndrome information, treatment principles, matched herbs, missed herbs, extra herbs, and H2 path support can be inspected together. This view helps identify whether an error arises from syndrome prediction, treatment mapping, H2 herb scoring, or dose estimation.

\subsection{Clinical Validation on Real CAP Cases}

We further evaluated the trained model on 50 real-world community-acquired pneumonia (CAP) cases collected with a clinical research form. This validation set was used as an external clinical check rather than as training data. Table~\ref{tab:clinical-validation} reports automatic metrics, prescription-level correctness checks, and physician audit results. In Step 1, the model reached 83.7\% accuracy, 31.88\% Macro-F1, and 92.8\% top-3 accuracy. In standalone Step 2 prescription prediction, Herb-F1 was 33.8\%, Jaccard was 0.2359, and matched-herb dose MAE was 3.541. In the connected Step 1-to-Step 2 setting, Herb-F1 was 31.57\%, Jaccard was 20.47\%, and dose MAE was 3.8374.

\begin{figure*}[t]
  \centering
  \includegraphics[width=0.86\textwidth,height=0.28\textheight,keepaspectratio]{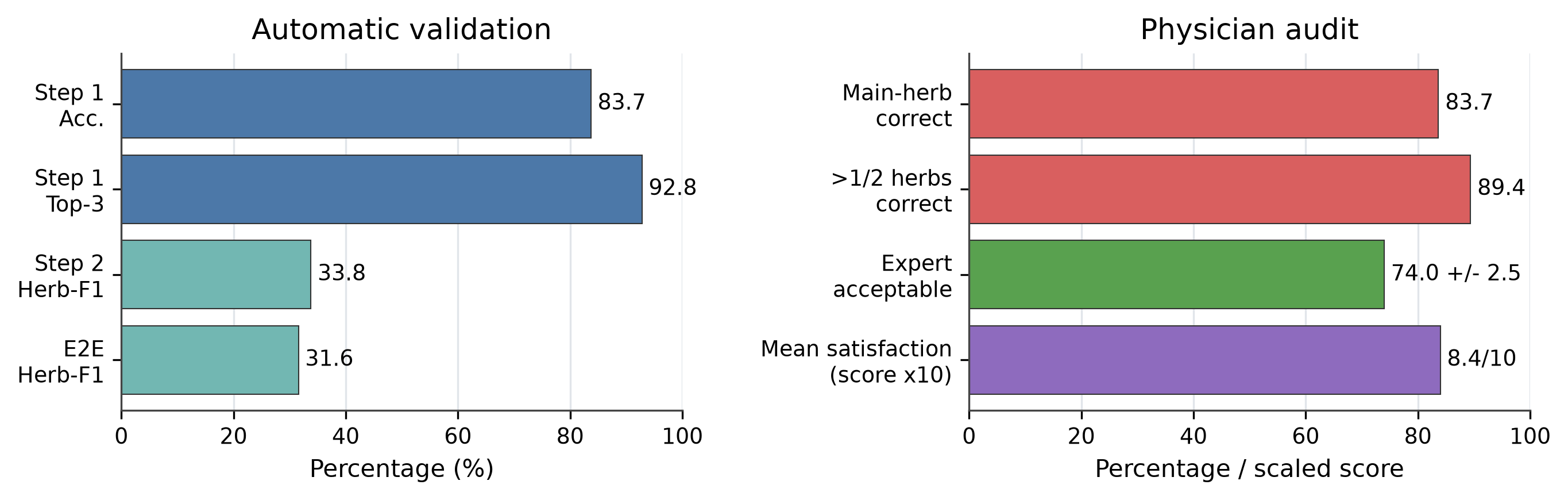}
  \caption{Simplified clinical validation summary on 50 real-world CAP cases and physician audit results from 10 TCM physicians. The figure reports core automatic validation metrics and prescription-level review outcomes.}
  \label{fig:clinical-validation}
\end{figure*}

We also used two prescription-level clinical criteria. When the primary herbs were correct and the prescription decision was judged correct, the accuracy was 83.67\%. When more than half of the herbs were correct and the prescription decision was judged correct, the accuracy was 89.40\%. Ten TCM physicians then reviewed the model-generated prescriptions. They rated 74.0\(\pm\)2.5\% of prescriptions as pharmacologically correct and acceptable. The mean satisfaction scores were 8.0/10 for prescription consistency, 8.6/10 for prescription rationality, and 8.6/10 for prescription usability. Fig.~\ref{fig:clinical-validation} visualizes these validation and audit results.

These endpoints should be interpreted at different granularities. Herb-F1 is a strict set-overlap metric under exact herb matching and therefore penalizes clinically similar but nonidentical formulas. The prescription-level checks are coarser clinical acceptability criteria based on primary-herb correctness or majority-herb correctness together with physician judgment. They are not substitutes for Herb-F1, and their higher values should not be read as evidence that all individual herbs and doses are correct. Additional reporting details and metric definitions are provided in the supplementary material.

\begin{table*}[t]
\centering
\caption{Clinical Validation on 50 Real-World CAP Cases}
\label{tab:clinical-validation}
\begin{tabular}{lll}
\toprule
Evaluation & Metric & Result \\
\midrule
Step 1 syndrome differentiation & Accuracy / Macro-F1 / Top-3 Acc. & 83.7\% / 31.88\% / 92.8\% \\
Standalone Step 2 prescription & Herb-F1 / Jaccard / Dose MAE & 33.8\% / 0.2359 / 3.541 \\
Connected Step 1-to-Step 2 & Herb-F1 / Jaccard / Dose MAE & 31.57\% / 20.47\% / 3.8374 \\
Prescription-level check & Primary herb correct and prescription correct & 83.67\% \\
Prescription-level check & More than half herbs correct and prescription correct & 89.40\% \\
Physician audit (10 TCM physicians) & Pharmacologically correct and acceptable & 74.0\(\pm\)2.5\% \\
Physician audit (10 TCM physicians) & Consistency / rationality / usability score & 8.0 / 8.6 / 8.6 out of 10 \\
\bottomrule
\end{tabular}
\end{table*}

\section{Discussion and Limitations}

The experiments support a conservative conclusion: in low-resource TCM prescription modelling, structured and inspectable constraints are more reliable than unconstrained language-model prescription generation. Direct Qwen2.5 generation remains weaker than structured H2 inference in herb-set alignment (Table~\ref{tab:step2}), and the no-H2-path ablation shows that hypergraph path features contribute beyond patient-only scoring (Table~\ref{tab:ablation}). The main evidence for the framework is therefore not a large absolute Herb-F1 margin, but the combination of improved syndrome differentiation, competitive prescription prediction, and auditable patient-conditioned reasoning paths.

The current results should not be overstated. TCM-BEST4SDT contains only 300 prescription cases, and the gain of patient-conditioned residual H2 over learned fixed H2 and strong constrained LLM baselines is small. The 50-case CAP evaluation and 10-physician audit provide a useful retrospective clinical check, but they are not evidence of prospective clinical efficacy. Dose MAE is also computed only on matched herbs, so it does not evaluate complete prescription-dose safety, contraindications, or herb interactions.

The end-to-end results highlight a remaining bottleneck: prescription performance degrades when treatment principles must be inferred or approximated from the predicted syndrome. The mapped-treatment setting drops to 0.2550 Herb-F1, indicating that syndrome-to-treatment alignment is fragile. Future work should normalize syndrome and treatment-principle concepts across datasets, use clinician-reviewed synonym mappings, evaluate semantic agreement, and conduct prospective dose-safety and clinical review before deployment claims are made.

\section{Conclusion}

We presented a patient-conditioned dual hypergraph framework for auditable TCM prescription support. The framework uses H1 for symptom--syndrome--treatment reasoning and H2 for treatment--formula--herb--dose reasoning. Language models serve as text-processing components rather than unconstrained prescribers. The core mechanism is dynamic hypergraph weighting: patient representations modulate H1 hyperedge incidences and H2 path weights while fixed TCM priors remain as structural anchors. Final experiments on TCM-SD, TCM-BEST4SDT, and 50 real-world CAP validation cases show that dynamic H1 improves syndrome differentiation with a strong encoder, dynamic H2 remains competitive against retrieval, static hypergraph, and constrained LLM baselines, and the full pipeline approaches oracle prescription performance. These results support patient-conditioned dual hypergraph reasoning as a useful structure for auditable TCM decision support. They also indicate the need for stronger consistency-loss validation, dose-safety review, and prospective clinical evaluation before deployment claims can be made.

\bibliographystyle{IEEEtran}
\bibliography{references}

\end{document}